\documentclass[9pt,conference]{IEEEtran}
\usepackage{amssymb,amsthm,amsmath,array}
\usepackage{graphicx}
\usepackage[caption=false,font=footnotesize]{subfig}
\usepackage{xspace}
\usepackage[sort&compress, numbers]{natbib}
\usepackage{stmaryrd}
\usepackage{xcolor}
\usepackage{mathtools}
\usepackage{float}
\usepackage{textcomp}

\usepackage{graphicx}
\usepackage[utf8]{inputenc}
\usepackage{lipsum}
\usepackage[hidelinks]{hyperref}
\usepackage{xspace}
\usepackage{bbm}
\usepackage{manfnt}
\usepackage{url}
\usepackage{xcolor}
\usepackage{soul}
\usepackage[british]{babel}
\usepackage{wrapfig}
\usepackage[font=scriptsize]{caption}

\DeclareMathOperator*{\sign}{sign}
\newcommand{\scp}[3][]{#1\langle #2, #3 #1\rangle}
\newcommand{\rop}{\cl A^{\rm v}}

\newcommand{\drop}{\cl B}
\newcommand{\ie}{\emph{i.e.}, }
\newcommand{\eg}{\emph{e.g.}, }
\newcommand{\sopu}{{\scriptscriptstyle\rm opu}}

\newcommand{\edt}[1]{{\color{black} #1}}

\newcommand{\bs}{\boldsymbol}
\newcommand{\bb}{\mathbb}
\newcommand{\cl}{\mathcal}
\newcommand{\ts}{\textstyle}

\newcommand{\iid}{%
  \ifmmode
  \mathrm{i.i.d.}%
  \else%
  i.i.d.\@\xspace%
  \fi%
}

\newcommand{\ignore}[1]{}

\newtheorem{theorem}{Theorem}[]

\begin{document}
\title{Signal processing after \\ quadratic random sketching with optical units}
\author{\IEEEauthorblockN{
        Rémi Delogne\IEEEauthorrefmark{1}, 
        Vincent Schellekens\IEEEauthorrefmark{2}, and 
        Laurent Daudet\IEEEauthorrefmark{3},
        Laurent Jacques\IEEEauthorrefmark{1}
    }
    \IEEEauthorblockA{
        \IEEEauthorrefmark{1} ICTEAM, UCLouvain, Belgium.\\
        \IEEEauthorrefmark{2} CEA, Paris.\\
        \IEEEauthorrefmark{3} LightOn, Paris.}
}
\maketitle
\begin{abstract}
    Random data sketching (or projection) is now a classical technique enabling, for instance, approximate numerical linear algebra and machine learning algorithms with reduced computational complexity and memory. In this context, the possibility of performing data processing (such as pattern detection or classification) directly in the sketched domain without accessing the original data was previously achieved for linear random sketching methods and compressive sensing. In this work, we show how to estimate simple signal processing tasks (such as deducing local variations in a image) directly using random quadratic projections achieved by an optical processing unit. The same approach allows for naive data classification methods directly operated in the sketched do- main. We report several experiments confirming the power of our approach.
\end{abstract}

\section{Introduction}\label{section:intro}
In machine learning and related areas, \textit{random sketching} techniques are now common to relieve computational costs of certain methods \cite{achlioptas_database-friendly_2001,rahimi_random_nodate,baraniuk_simple_2008}. While introducing randomness in the data, such techniques allow for controlled probabilistic bounds on approximation errors committed in the process. 

In order to work with sketched data, one will often require to access the sketches of functions of the original data rather than simply sketches obtains by randomly projecting original data. For example one can think of a video stream where only parts of the frames will require attention, requiring means of restricting the images to a particular area. While this may be feasible when the sketching is essentially linear (\cite{SPWCM}), this is no easy task with non-linear sketching.

In this work we will restrict ourselves to a particular non-linear type of sketching called \textit{quadratic random sketching}. This transformation is inspired by recent Optical Processing Units (OPU) capable of simulating random projections at the speed of light and at very low power (\cite{ohana_kernel_2020}). We show that estimating functions of the output of random projections is possible. 

\section{Rank -one random projections of signals}\label{section:maths}

This section presents the mathematical framework of \textit{quadratic random sketching}. This transformation relies on the \textit{rank-one projection} (ROP) of the observed signal and it is crucial to understand this before getting to the description of the OPU.

{Quadratic random sketching} consists in taking a series of $m$ measurements $(\bs a_i^\top\bs x)^2$ of a signal of interest $\bs x\in\bb R^n$, with a set of $m$ random vectors $\{\bs a_i\}_{i=1}^m \subset \bb R^n$. The sketching operator $\rop$ is defined as
\begin{equation}
\label{eq:originalROP}
\rop: \bs x \in \bb R^n \mapsto \rop(\bs x) := \big( (\bs a_i^\top\bs x)^2 \big)_{i=1}^m \in \bb R^m_+,
\end{equation}
which as observed in the context of phase retrieval~\cite{phaselift}, amounts to a ROP of the \emph{lifted signal}, \ie the rank-one matrix $\bs X = \bs x \bs x^\top \in \bb R^{n \times n}$, onto the rank-one random matrices $\{\bs A_i := \bs a_i \bs a_i^\top\}_{i=1}^m \subset \bb R^{n \times n}$. We thus use ``quadratic sketch'' and ``ROP measurements'' interchangeably.   


As we will see later, an OPU allows us to compute all the components of $\rop(\bs x)$ in a reproducible way using the physical properties of multiple scattering of coherent light in random media. In this context, the vectors $\{\bs a_i\}_{i=1}^m$ are fixed and Gaussian (\cite{saade_random_2016}), but \emph{hidden} to us. 

The ROP operator is sadly \textit{biased} (non-isotropic). In other words, the expectation of the squared norm of the ROP of a vector $\bs x$ is not proportional to the squared Frobenius norm of the lifted signal $\bs x\bs x^\top$ (equal to the fourth power of the norm of $\bs x$) \cite{chen_exact_2015}. \edt{This is relatively obvious considering that the ROP operator only outputs positive values.} However, isotropy is a key property to ensure that the ROP sketch retains essential information on $\bs x$. This leads to the definition of a debiased ROP operator (DROP) which rids us of the bias (see \cite{chen_exact_2015}, lemma 4 and appendix F):
\begin{equation}
\label{eq:DROP}
\drop: \bs x \in \bb R^{n} \mapsto \drop(\bs x)= \big( \rop_{2i}(\bs x) - \rop_{2i+1}(\bs x) \big)_{i=1}^{m}.
\end{equation}

This new debiased estimator DROP can easily be implemented on an OPU by first applying the operator $\rop$ on a vector then splitting it in two and subtracting one half from the other. Thanks to the constant computational complexity of $\rop$ on the OPU, this has little impact on the total computing time.

\section{The Optical Processing Unit}\label{section:opu}

The OPU itself works by encoding a binary array onto a laser using an array of micro-mirrors. The resulting beam is then sent through a scattering medium that acts as the desired quadratic sketch. The OPU process thus requires a binarisation step. 

When grappling with OPUs one needs to carefully study the particulars of the transformation it applies. As is the case with most physical systems it falls short from exactly replicating the theoretical model. A careful analysis (see \cite{icassp23}) shows that with a binary input array $\bs x \in \{0,1\}^n$ (with $n$ up to $O(10^6)$), the OPU will roughly output
\begin{equation}
\label{eq:opu-noisy-sensing}
y = \cl Q_b( |\scp{\bs a}{\bs x}|^2 + \edt \eta),
\end{equation}
with the uniform quantiser $\cl Q_b(t)$ equal to $\delta \lfloor t/\delta \rfloor$ if $0\leq t\leq C$, and $C$ otherwise, the bin width $\delta = C 2^{-b}$, the bit depth $b=8$, and a noise $\edt \eta$ such that $|\edt \eta|\leq 4 \delta$ with high probability, provided that the sparsity of the input vector remains between $20\%$ and $80\%$. In this case, the OPU's behaviour is tame enough to use it as the theory predicts. 

\begin{figure}[t]
\centering
\includegraphics[width=8.5cm]{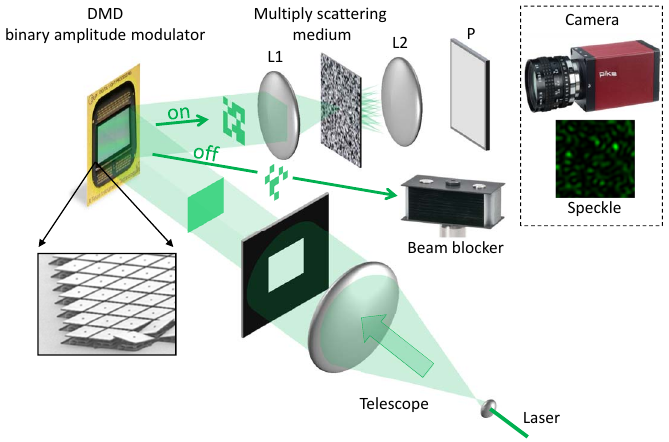}
\caption{\cite{saade_random_2016} The OPU takes a laser and uses an array of mirrors to encode binary data in the beam \cite{liutkus2014imaging}. It then sends the beam through a scattering medium replicating the effect of the operator $\rop$ on the signal encoded in the light beam.}\medskip
\label{fig:opu}
\end{figure}

\section{Processing in the sketched domain}

In the heart of our method, lies the \textit{Sign Product Embedding}. This proposition shows that $\scp{\sign(\drop(\bs u))}{\drop(\bs x)}$ can act as a proxy for $|\scp{\bs u}{\bs x}|^2$. A wise choice of $\bs u$ could hence give us access to local information about $\bs x$ for example, using only the sketch $\drop (\bs x)$ and $\drop(\bs u)$ without ever reconstructing $\bs x$, thus solving the problem described in the introduction. This is more formally stated in the following theorem.

\begin{theorem}[Sign Product
Embedding of DROP Sketches]
\label{prop:drop-spe}
Given a fixed unit vector $\bs u \in \bb R^n$, $\kappa = \pi/4$, and a distortion $0<\delta <1$, provided that 
\begin{equation}
\label{eq:sample-complex-drop-spe}
\ts \ts m \geq C \delta^{-2} k \log(\frac{n}{k\delta}),    
\end{equation}
then, with probability exceeding $1 - C \exp(-c \delta ^2 m)$, for all $k$-sparse signals $\bs x \in \Sigma_k := \{\bs v \in \bb R^n: |{\rm supp}(\bs v)|\leq k\}$, $\cl B$ respects the SPE over $\Sigma_k$, \ie 
\begin{equation}
\label{eq:drop-spe}
\ts \Big|\frac{\kappa}{m} \scp{\sign(\drop(\bs u))}{\drop(\bs x)} - {\scp{\bs u}{\bs x}^2}\ignore{{\|\bs u\|^2}} \Big| \leq \delta \|\bs x\|^2,    
\end{equation}    
with $\sign$ the sign operator applied componentwise on vectors. 
\end{theorem}

This theorem, proved in \cite{esann22} using a covering argument, shows that if $m$ is large enough, the quantity $\frac{\kappa}{m} \scp{\sign(\drop(\bs u))}{\drop(\bs x)}$ approximates $\scp{\bs u}{\bs x}^2$ with a controlled distortion $\delta$ scaling like $O\big(\sqrt{k/m}\,\big)$ up to log factors. It is therefore possible to approximate linear functions of any signal $\bs x\in\bb R^n$, by projecting their sketches $\drop(\bs x)$ on $\sign(\cl B(\bs u)) \in \{\pm 1\}^m$.

\section{Experiments}

\begin{figure}[t]
\centering
\raisebox{5mm}{\includegraphics[width=.34\columnwidth]{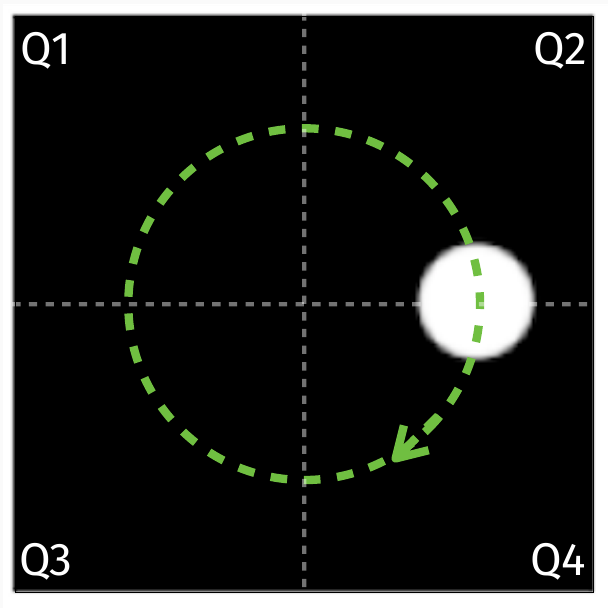}}
\includegraphics[width=.64\columnwidth]{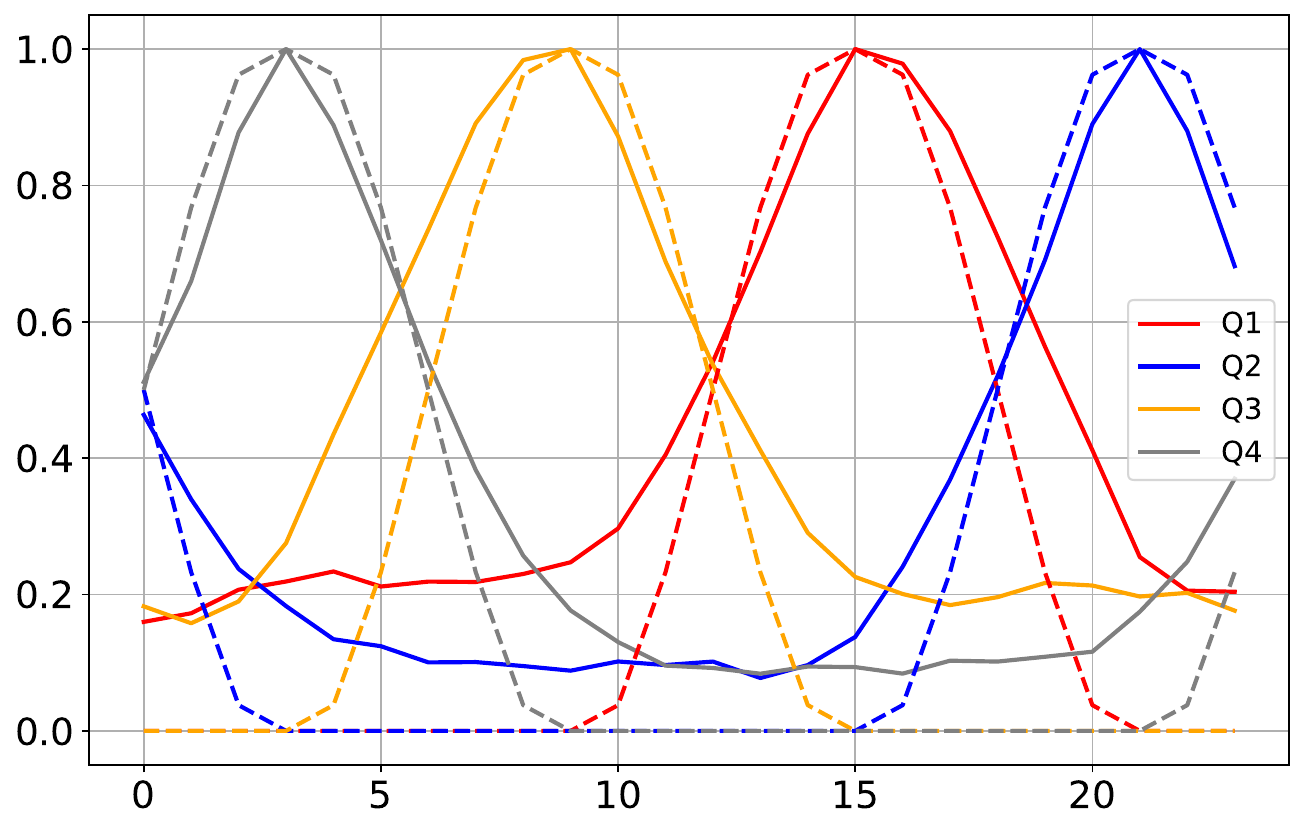}
\caption{(Left) A frame from the synthetic video of a white rotating disk on a black background. (Right) Evolution of the true and estimated quadrant occupancies, in dashed and plain curves, respectively}
\label{fig:disk}
\end{figure}

In this section we demonstrate two applications of our method with two simple classification experiments. First we consider a synthetic video consisting of 24 vectorised $950\times 950$ binary images $\{\bs x_t\}_{t=0}^{23}$, representing a rotating disk (white) on a black (zero) background (see Fig.~\ref{fig:disk} (left)). Our objective is to detect the passage of the disk in each of the four quadrants of the image solely based on the $m-$dimensional OPU measurements $\{\drop_{\sopu}(\bs x_t)\}_{t=0}^{23}$. 

We define four quadrant indicators, \ie four vectors $\bs u_i\in\bb R^n$ equal to 1 in the $j$-th quadrant and zero outside ($j\in\{1,\ldots,4\}$). Thanks to Thm~\ref{prop:drop-spe}, we can estimate, up to some distortion, the $j$-th \emph{quadrant occupancy} signal $q_j(t) := |\scp{\bs u_j}{\bs x_t}|^2$ from the estimated occupancy 
\begin{multline*}
\ts q^{\rm est}_{j,\sopu}(t) := \frac{\kappa}{m} \scp{\sign(\drop_{\sopu}(\bs u))}{\drop_{\sopu}(\bs x_t)}\\ 
\ts \simeq q^{\rm est}_j(t) := \frac{\kappa}{m} \scp{\sign(\drop(\bs u))}{\drop(\bs x_t)}.
\end{multline*}

The curves (normalised to their maximum values) in Fig.~\ref{fig:disk} (right) show the evolution with time of $q^{\rm est}_{j,\sopu}(t)$ with $m=10\,000$. A quick comparison with the true curves (dashed) shows that we can correctly estimate quadrant occupancy based solely on the sketches of the full original frames.


Our second experiment consists in a simple classification task on a binarised version of the handwritten digits of the MNIST dataset $\cl X := \{(\bs x_k, t_k)\}_{k=1}^{N=70\,000} \subset \{0,1\}^{n=28^2} \times \{0,\ldots,9\}$~\cite{mnist}. Here, we aim to compare the results of a naive classification performed in the direct domain, to a classification operated in the sketched domain. We start by randomly splitting $\cl X$ into a training $\cl X_{\rm tr}$ and a test set $\cl X_{\rm te}$ according to a split of $60\,000$ and $10\,000$ images, respectively. From $\cl X_{\rm tr}$ we compute the 10 centroids $\{\bs c_j\}_{j=0}^9 \subset \bb R^{n}$ of each class of digits and we define 10 vectors $\bs u_j = \bs c_j /\|\bs c_j\|$. In the direct domain and for images of the test set, the estimated label of an image $\bs x_k$ is then defined as $\hat{t}_k := \arg\max_j |\scp{\bs u_j}{\bs x_k}|^2$. 



Next we compute the sketches of the centroids and of the remaining instances with the OPU, and the label estimate of a test image $\bs x_k$ is then computed by comparing the two transformations 
$$
\ts \hat{t}^{\rm sk}_k := \arg\max_j \frac{\kappa}{m} \scp{\sign(\drop_{\edt\sopu}(\bs u_j))}{\drop_{\edt\sopu}(\bs x_k)},
$$
which should be close to $\hat{t}_k$ according to Thm~\ref{prop:drop-spe}. We summarise the different average testing accuracies reached for both the direct and the sketched classifications in Table~\ref{tab:MNIST-table}. Keeping in mind that our classification method is rudimentary (as shown in, \eg \cite{saade_random_2016}, it is possible to develop much better, non-linear classification algorithms directly in the sketched domain), we observe anyway that as $m$ increases the average accuracy improves and approaches the one performed in the direct domain. The loss in accuracy can also be attributed to several other factors such as the binarisation process and noise within the OPU.  

\begin{table}[t]
    \centering
\scriptsize
\begin{tabular}{c|cccccc}
  & Direct & $m=200$ 
  & $400$ & $800$ & $1600$&$3200$
  \\ 
  &&&&&&\\[-2.5mm]
  \hline
  &&&&&&\\[-2mm]
 Accuracy  $[\%]$ & $82.1$ & $56.3$
 & $59.2$ & $66.9$&$71.8$&$75.0$
 \end{tabular}
    \caption{Testing accuracy (for the second experiment) (in $\%$) in the direct domain (without sketching), and in the sketched domain for various values of $m$.}
    \label{tab:MNIST-table}
\end{table}

\section{Conclusion}

We have shown that signal estimation in the sketched domain is possible with the optical quadratic random sketching delivered by an OPU using the \emph{sign product embedding}. We illustrated it with two toy examples to demonstrate the possibility of extracting localised information from a sketched signal and classifying images from their sketches.

Future works could exploit both the OPU calibration and the dependence of the SPE distortion in the sketch dimension to formalise precise statistical tests for pattern matching in the sketched domain. On a more theoretical note, preliminary numerical tests show that the SPE of the DROP could hold for matrices of rank greater than 1. This is appealing for, \eg change point detection application in a data stream $\{\bs x_t\}_{t \in \bb Z}$ since, for instance, a time change in ROP sketches $\rop(\bs x_{t+1}) - \rop(\bs x_{t})$ is equivalent to the ROP of the rank-2 matrix $\bs x_{t+1}\bs x_{t+1}^\top - \bs x_{t}\bs x_{t}^\top$.
\bibliographystyle{IEEEtran}
\bibliography{references}

\begin{thebibliography}{10}
\providecommand{\url}[1]{#1}
\csname url@samestyle\endcsname
\providecommand{\newblock}{\relax}
\providecommand{\bibinfo}[2]{#2}
\providecommand{\BIBentrySTDinterwordspacing}{\spaceskip=0pt\relax}
\providecommand{\BIBentryALTinterwordstretchfactor}{4}
\providecommand{\BIBentryALTinterwordspacing}{\spaceskip=\fontdimen2\font plus
\BIBentryALTinterwordstretchfactor\fontdimen3\font minus
  \fontdimen4\font\relax}
\providecommand{\BIBforeignlanguage}[2]{{%
\expandafter\ifx\csname l@#1\endcsname\relax
\typeout{** WARNING: IEEEtran.bst: No hyphenation pattern has been}%
\typeout{** loaded for the language `#1'. Using the pattern for}%
\typeout{** the default language instead.}%
\else
\language=\csname l@#1\endcsname
\fi
#2}}
\providecommand{\BIBdecl}{\relax}
\BIBdecl

\bibitem{achlioptas_database-friendly_2001}
D.~Achlioptas, ``Database-friendly random projections,'' in \emph{Proceedings
  of the twentieth ACM SIGMOD-SIGACT-SIGART symposium on Principles of database
  systems}, 2001, pp. 274--281.

\bibitem{rahimi_random_nodate}
A.~Rahimi and B.~Recht, ``Random features for large-scale kernel machines,'' in
  \emph{Advances in Neural Information Processing Systems}, J.~Platt,
  D.~Koller, Y.~Singer, and S.~Roweis, Eds., vol.~20.\hskip 1em plus 0.5em
  minus 0.4em\relax Curran Associates, Inc., 2007.

\bibitem{baraniuk_simple_2008}
R.~Baraniuk, M.~Davenport, R.~DeVore, and M.~Wakin, ``A simple proof of the
  restricted isometry property for random matrices,'' \emph{Constructive
  Approximation}, vol.~28, no.~3, pp. 253--263, 2008.

\bibitem{SPWCM}
M.~A. Davenport, P.~T. Boufounos, M.~B. Wakin, and R.~G. Baraniuk, ``Signal
  processing with compressive measurements,'' \emph{IEEE Journal of Selected
  topics in Signal processing}, vol.~4, no.~2, pp. 445--460, 2010.

\bibitem{ohana_kernel_2020}
R.~Ohana, J.~Wacker, J.~Dong, S.~Marmin, F.~Krzakala, M.~Filippone, and
  L.~Daudet, ``Kernel computations from large-scale random features obtained by
  optical processing units,'' in \emph{ICASSP 2020-2020 IEEE International
  Conference on Acoustics, Speech and Signal Processing (ICASSP)}.\hskip 1em
  plus 0.5em minus 0.4em\relax IEEE, 2020, pp. 9294--9298.

\bibitem{phaselift}
E.~Candès, T.~Strohmer, and V.~Voroninski, ``Phaselift: Exact and stable
  signal recovery from magnitude measurements via convex programming,''
  \emph{Comm. Pure Appl. Math.}, pp. 66: 1241--1274, 2013.

\bibitem{saade_random_2016}
A.~Saade, F.~Caltagirone, I.~Carron, L.~Daudet, A.~Dr{\'e}meau, S.~Gigan, and
  F.~Krzakala, ``Random projections through multiple optical scattering:
  Approximating kernels at the speed of light,'' in \emph{2016 IEEE
  International Conference on Acoustics, Speech and Signal Processing
  (ICASSP)}.\hskip 1em plus 0.5em minus 0.4em\relax IEEE, 2016, pp. 6215--6219.

\bibitem{chen_exact_2015}
Y.~Chen, Y.~Chi, and A.~J. Goldsmith, ``Exact and stable covariance estimation
  from quadratic sampling via convex programming,'' \emph{IEEE Transactions on
  Information Theory}, vol.~61, no.~7, pp. 4034--4059, 2015.

\bibitem{icassp23}
\BIBentryALTinterwordspacing
D.~R., D.~L. Schellekens~V., and J.~L., ``Signal processing with quadratic
  optical sketches,'' \emph{Intarnational conference on acoustics and signal
  processing 2023}, 2022. [Online]. Available:
  \url{https://arxiv.org/abs/2212.00660}
\BIBentrySTDinterwordspacing

\bibitem{liutkus2014imaging}
A.~Liutkus, D.~Martina, S.~Popoff, G.~Chardon, O.~Katz, G.~Lerosey, S.~Gigan,
  L.~Daudet, and I.~Carron, ``Imaging with nature: Compressive imaging using a
  multiply scattering medium,'' \emph{Scientific reports}, vol.~4, no.~1, pp.
  1--7, 2014.

\bibitem{esann22}
\BIBentryALTinterwordspacing
R.~Delogne, V.~Schellekens, and L.~Jacques, ``{ROP} inception: signal
  estimation with quadratic random sketching,'' \emph{European Symposium on
  Artificial Neural Networks}, 2022. [Online]. Available:
  \url{https://arxiv.org/abs/2205.08225}
\BIBentrySTDinterwordspacing

\bibitem{mnist}
L.~Deng, ``The {MNIST} database of handwritten digit images for machine
  learning research [best of the web],'' \emph{IEEE signal processing
  magazine}, pp. vol. 29, no 6, p. 141--142, 2012.

\end{thebibliography}
\end{document}